\documentclass[12pt,preprint]{aastex}
\begin{document}
\newcommand{\calu}{{\cal U}}
\newcommand{\calq}{{\cal Q}}
\newcommand{\bx}{{\rm \bf x}}
\newcommand{\bk}{{\bar{\kappa}}}
\title{Optimal Weak Lensing Skewness Measurements}
\author{Tong-Jie Zhang}
\affil{Department of Astronomy, Beijing Normal University, Beijing 100875,
P.R.China;
tjzhang@bnu.edu.cn; and
Canadian Institute for Theoretical Astrophysics, University of Toronto,M5S
3H8,Canada;
tzhang@cita.utoronto.ca}
\author{Ue-Li Pen}
\affil{Canadian Institute for Theoretical Astrophysics, University of
Toronto, M5S 3H8, Canada; pen@cita.utoronto.ca}
\author{Pengjie Zhang}
\affil{NASA/Fermilab Astrophysics Group,
Fermi National Accelerator Laboratory,
Box 500, Batavia, IL 60510-0500; zhangpj@fnal.gov }
\author{John Dubinski}
\affil{Canadian Institute for Theoretical Astrophysics, University of
Toronto, M5S 3H8, Canada, dubinski@cita.utoronto.ca}

\begin{abstract}

Weak lensing measurements are starting to provide
statistical maps of the distribution of matter in the universe
that are increasingly precise and complementary to cosmic microwave
background maps.
The most common measurement is the correlation in alignments of
background galaxies which can be used to infer the variance 
of the projected surface density of matter. This measurement of
the fluctuations is insensitive to the total mass content and is
analogous to using waves on the ocean to measure its depths.  
However, when the depth is shallow as happens near a beach waves 
become skewed.  Similarly, a measurement of skewness in the projected matter 
distribution directly measures the total matter content of the universe.  
While skewness has
already been convincingly detected, its constraint on cosmology is
still weak.  We address optimal analyses for the CFHT Legacy Survey in
the presence of noise.  We show that a compensated Gaussian filter
with a width of 2.5 arc minutes optimizes the cosmological
constraint, yielding $\Delta \Omega_m/\Omega_m\sim 10\%$. This is
significantly better than other filters which have been considered in
the literature.  This can be further improved with tomography and other
sophisticated analyses.

\end{abstract}


\section{Introduction}

Mapping the mass distribution of matter in the Universe has
been a major challenge and focus of modern observational
cosmology.  The only direct procedure to weigh the matter
in the universe is by using the deflection of light by gravity.  
While this effect is very small, a large statistical sample can provide a
precise measurement of averaged quantities.

There are very few direct ways to weigh the universe.  The most
accurate measurement by combining CMB data with large scale
structure \citep{2003astro.ph..2209S,2003astro.ph..2435C} results
in $\Omega_m\sim 0.27$ with zero geometric curvature implying
a cosmological constant $\Omega_\Lambda \sim 0.73$.  This
type of inference requires combining data measured at
different times and on different length scales.
\citet{2003astro.ph..4237B} have shown that the same data can be
consistent with $\Omega_m=1$ if one gives up perfect scale
invariance for the primordial perturbations and allows for a
neutrino mass of 1eV.  The physical constraint arises since
the CMB measures the fluctuations on large scales $L\sim $Gpc at
high redshift $z \sim 1100$.  The large-scale structure measures
scales of $L\sim 1-100$ Mpc and a low redshift of $z\sim 0$.  The
scales only have a small overlap \citep{2002PhRvD..66j3508T}. If
one requires perfect scale invariance of the fluctuations, one is
forced into a low matter density with a cosmological constant.  It
is perhaps an aesthetic choice to trade scale invariance in time
to scale invariance in space.

Weak gravitational lensing provides a direct statistical
measure of the dark matter distribution regardless of the nature and
dynamics of both the dark and luminous matter intervening between the
distant sources and observer.  Weak lensing by
large-scale structure can lead to the shear and magnification of the
images of distant faint galaxies. Based on the theoretical work done
by \cite{1967ApJ...147...61G}, \cite{1991MNRAS.251..600B},
\cite{1991ApJ...380....1M} and \cite{1992ApJ...388..272K} performed
the first calculation of weak lensing by large-scale structure, the
result of which showed the expected distortion amplitude of weak
lensing effect is at a level of roughly a few percent in adiabatic
cold dark matter models. \cite{1992ApJ...388..272K} also made early
predictions for the power spectrum of the shear and convergence using
linear perturbation theory.  Due to the weakness of the effect, all
detections have been statistical in nature, primarily in regimes where
the signal-to-noise is less than unity.  Fortunately, several groups
have been able to measure this weak gravitational lensing effect
\citep{2000MNRAS.318..625B,2002ApJ...572L.131R,2002ApJ...572...55H,
2002A&A...393..369V,2002astro.ph.10604J,2002astro.ph.10213B,
2002astro.ph.10450H}recently.

In the standard model of cosmology, fluctuations start off small,
symmetric and Gaussian.  Even in some non-Gaussian models like
topological defects, initial fluctuations are still symmetric:
positive and negative fluctuations occur with equal probability
\citep{1994PhRvD..49..692P}.  As fluctuations grow by gravitational
instability, this symmetry can no longer be maintained - over densities
can be arbitrarily large, while under dense regions can never have
less than zero mass.  This leads to a skewness in the distribution of
matter fluctuations.  While skewness has already been measured at very
high statistical significance \citep{2003astro.ph..2031P}, the
measurement has not resulted in a strong constraint on the total
matter density $\Omega_m$.  The data has so far been limited by sample
variance and analysis techniques.  Currently ongoing surveys, such as
the Canada-France-Hawaii-Telescope(hereafter CFHT) Legacy Survey,
will provide more
than an order of magnitude improvement in the statistics.  In this
paper, we address the optimal analysis of the new data sets and
examine the likely plausible accuracies on the direct measurements of
matter density that they can achieve.  The calculations rely only on
sub-horizon Newtonian physics.

Several studies have addressed the feasibility of the skewness
measurements \citep{2000ApJ...530..547J,2000ApJ...537....1W}.  These
pioneering contributions provided the first estimates of the
expected strengths of the skewness $S_3$.  A real measurement is
limited by the sample variance in $S_3$ and noise properties.
Furthermore, the density field is always smoothed by some filters.
Since gravitational lensing can only measure differences in mass,
all such filters must have zero area.  In this paper we study a
range of filters that have been suggested in the literature.  Our
goal is to find the optimal scale for each filter, i.e. the scale
that maximizes the dependence on $\Omega_m$.  We study the
filters that have been mainly employed in the literature: top-hat,
Gaussian, Wiener, aperture and compensated Gaussian. Only the last
three have zero area, and can be applied to real data. For
completeness, we also compare the first two filters, on which much
of the literature is based.

In second order perturbation theory, one finds that the skewness
scales as the square of the variance and inversely to density. In
terms of the dimensionless surface density $\kappa$, one can
express the square of the variance and the skewness as
respectively $\langle \bk^2\rangle^{1/2} \propto
\sigma_8\Omega_m^{-0.75}$ and $S_3 \equiv \frac{\langle
\bk^3\rangle}{\langle \bk^2\rangle^2} \propto \Omega_m^{-0.8}$.
Therefore, one can break the degeneracy between $\sigma_8$ and
$\Omega_m$ if only both the variance and the skewness of the
convergence are measured.  The skewness of the convergence field
has been studied in perturbation theory
\citep{1997A&A...322....1B,1999ApJ...519L...9H} and initial detections have
been reported \citep{2002A&A...389L..28B}. \cite{2000ApJ...530..547J}
investigate weak lensing by large-scale structure using ray
tracing in N-body simulations. By smoothing the convergence field using a
top-hat window function, they compute $S_3$ under two conditions -
one with noise and one without noise added in the convergence fields 
by the third moment for all varieties of cosmological models. 
Moments are linear in the PDF: one can combine the moments of different
maps, which gives the same answer as combining maps first.  Non-linear
methods have also been proposed.
One can
measure $S_3$ is using the conditional
second moment of the $\kappa$ field, specifically, the second moment of
positive $\kappa$ and negative $\kappa$, which is related to $S_3$ in
perturbation theory 
\citep{1993ApJ...405..437N,1995ApJ...442...39J}. 




Moments are also additive in the presence of noise, such that skewness-free
noise (which realistic symmetric noise sources often have) does not bias
the measurement of moments.

\cite{2000ApJ...537....1W} presented a calculation for
the skewness $S_3$ and its standard deviation of weak lensing by
large scale-scale structure based on N-body simulations. By
smoothing the $\kappa$ field using a top-hat filter, they show
that the standard deviation of the skewness after adding simulated
shot noise to the $\kappa$ field are only slightly increased by about
16 per cent compared with the case of pure $\kappa$ field.

In this paper, we present the first extended comparison of
skewness for simulated weak lensing using different kinds of
window functions to isolate the filter that is optimal for
distinguishing cosmological models. We highlight some candidate
window functions that have been used separately in the
literature. The outline of the paper is as follows.  In \S 2, we
introduce the strategy of map construction of weak
lensing from simulations. In \S 3, we describe the detail of window functions employed
and present the results and summarize our conclusions
in Section 4.

\section{Simulations and map construction of weak lensing }
\subsection{N-body simulation}

We ran a series of N-body simulations with different values of
$\Omega_m$ to generate convergence maps and make simulated catalogs to
calibrate the observational data and estimate errors in the analysis.
The power spectra for given parameters were generated using CMBFAST
\citep{1996ApJ...469..437S} and these tabulated functions were used to
generate initial conditions.  The power spectra were normalized to be
consistent with the earlier two point analysis from this data set
\citep{2002A&A...393..369V}.  We ran all of the simulations using a
parallel, Particle-mesh N-body code (Dubinski, J., Kim, J., Park, C.
2003) at $1024^3$ mesh resolution using $512^3$
particles on an 8 node quad processor Itanium Beowulf cluster at CITA.
Output times were determined by the appropriate tiling of the light
cone volume with joined co-moving boxes from $z \approx 6$ to $z=0$.
We output periodic surface density maps at $2048^2$ resolution along
the 3 independent directions of the cube at each output interval.
These maps represent the raw output for the run and are used to
generate convergence maps in the thin lens and Born approximations by
stacking the images with the appropriate weights through the comoving
volume contained in the past light cone.

All simulations started at an initial redshift $z_i=50$, and ran for
1000 steps in equal expansion factor ratios with box size $L=200
h^{-1}$ Mpc comoving.  We adopted a Hubble constant $h=0.7$ and a
scale invariant $n=1$ initial power spectrum.  A flat cosmological
model with $\Omega_m + \Lambda = 1$ was used.  Four models were run,
with $\Omega_m$ of 0.2, 0.3, 0.4 and 1.  The power spectrum
normalization $\sigma_8$ was chosen as 1.16, 0.90, 0.82 and 0.57
respectively.

\subsection{Simulated Convergence Maps}
The convergence $\kappa$ is the projection of the matter over-density $\delta$
along the line of sight $\theta$ weighted by the lensing
geometry and source galaxy distribution. It can be expressed as
\begin{equation}
\kappa(\theta,\chi_s)=\int_0^{\chi_s} W(\chi) \delta(\chi,r(\chi)\theta))d\chi,
\end{equation}
where, $\chi$
is the comoving distance in unit of $c/H_0$, and $H_0=100\  h
\ {\rm km/s/Mpc}$. The weight
function $W(\chi)$ is\
\begin{equation}
W(\chi)=\frac{3}{2}\Omega_{m}g(\chi)(1+z)
\end{equation}
determined by  the source galaxy distribution function
$n(z)$ and the lensing geometry.
\begin{equation}
g(\chi)=r(\chi) \int_{\chi}^{\infty} d\chi' n(\chi')\frac{r(\chi'-\chi)}{r(\chi')}.
\label{eqn:gx}
\end{equation}
$r(\chi)$ is the radial coordinate. $r(\chi)=\sinh(\chi)$ for open,
$r(\chi)=\chi$ for flat and $r(\chi)=\sin(\chi)$ for closed geometry of
Universe. $n(z)=n(\chi)d\chi/dz$ is normalized such that $\int_0^{\infty} n(z)dz=1$.
For the CFHT Legacy Survey, we adopt
$n(z)=\frac{\beta}{z_0\Gamma(\frac{1+\alpha}{\beta})}(\frac{z}{z_0})^
\alpha\exp(-(\frac{z}{z_0})^\beta)$ with $\alpha=2$ and $\beta=1.2$ and
the source redshift parameter $z_0=0.44$, which peaks at $z_p=1.58z_0$, respectively.
The mean redshift is $\bar{z}=2.1z_0$ and the median redshift is
$z_h=1.9z_0$\citep{2002A&A...393..369V}. The source redshift distribution $n(z)$
adopted here is the same as that for VIRMOS.

\begin{figure}
\plotone{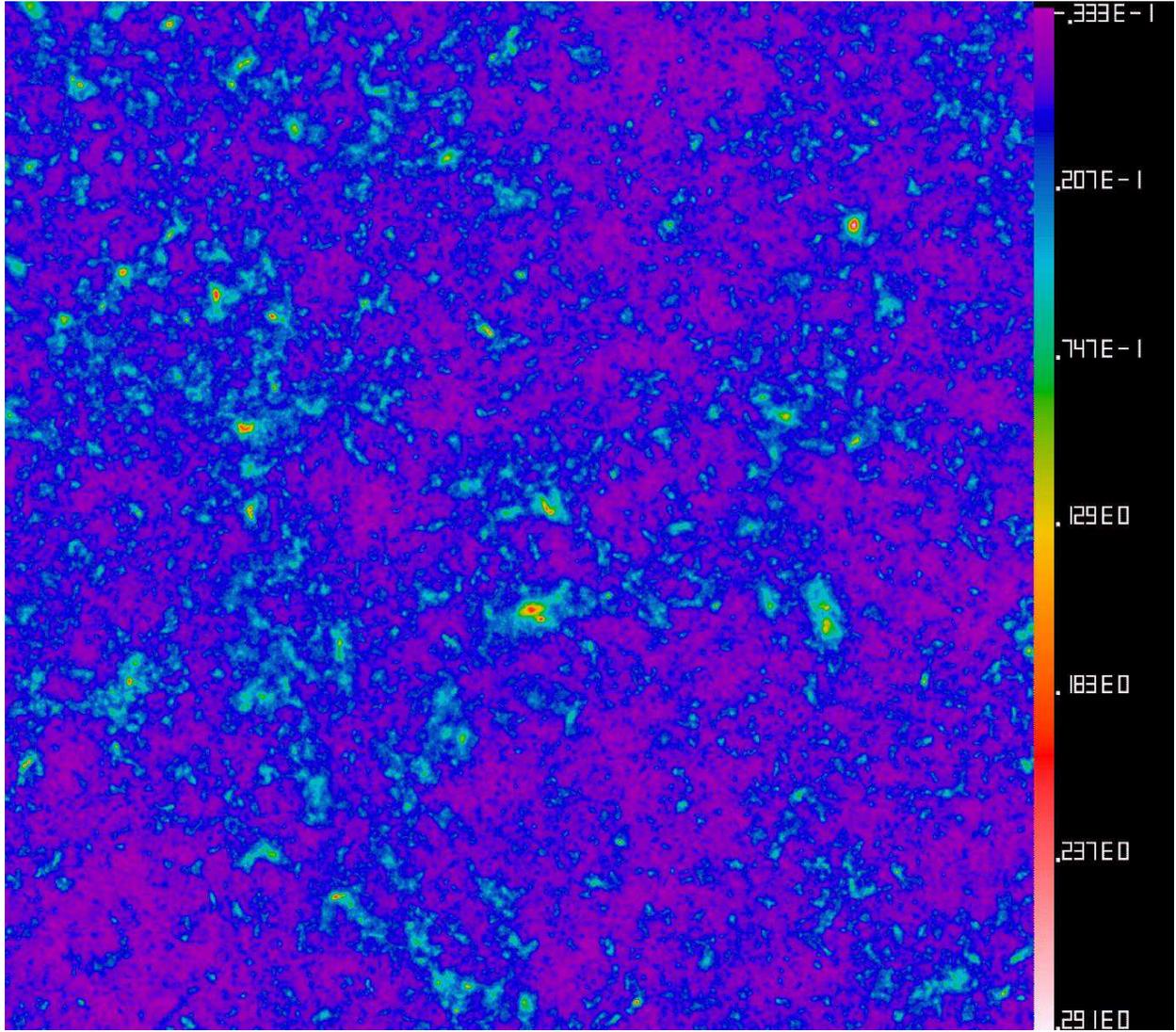}
\caption{An initial noise-free $\kappa$ map in the
N-body simulation of a $\Omega_m=$0.3 $\Lambda$CDM cosmology with
a map width of 3.02 degrees
and $2048^2$ pixels, and the scale is in units of $\kappa$.
}
\label{fig:kmap03}
\end{figure}

During each simulation we store 2D projections of $\delta$ through
the 3D box at every light crossing time through the box along all
x, y and z directions.  Our 2D surface density sectional maps are
stored on $2048^2$ grids. After the simulation, we stack sectional
maps separated by a width of the simulation box, randomly choosing
the center of each section and randomly rotating and flipping each
section.  The periodic boundary condition guarantees that there is
no discontinuities between any two adjacent boxes. We then add
these sections with the weights given by $W(z)$ onto a map of
constant angular size, which is generally determined by the
maximum projection redshift.  To minimize the repetition of the
same structures in the projection, we alternatively choose the
sectional maps of x, y, z directions during the stacking. Using
different random seeds for the alignments and rotations, we make
$40$ maps for each cosmological model. Since the galaxy
distribution peaks at $z\sim 1$, the peak contribution of lensing
comes from $z\sim 0.5$ due to the lensing geometry term.  Thus the
maximum projection redshift $z\sim 2$ is sufficient for the
lensing analysis. So we project the $\Omega_0=1,0.4,0.3$
simulations to $z=2$ and obtain $40$ maps each with angular width
$\theta_\kappa=4.09, 3.18$ and $3.02$ degrees, respectively. To
make sufficiently large maps, for $\Omega_0=0.2$, we project up to
$z=1.8$ and obtain maps with angular width $\theta_\kappa=2.86$
degrees. One $\kappa$ map created from a cosmological simulation
of $\Omega_m=$0.3 is shown in Fig.\ref{fig:kmap03}. The skewness
is quite apparent at this resolution of the simulation. Decreasing the
cosmological density while maintaining the same variance of
convergence $\kappa$ forces structures to be more non-linear, and
thus more skewed.  Our challenge is to extract this behavior
accurately from realistic data.

We then simulate the CFHT Legacy Survey by adding noise to these clean maps.
The noise $\kappa$ maps have  a pixel-pixel variance
$\sigma_N^2=\langle e^2\rangle/2/\langle N_{\rm pixel} \rangle$.
Here $\langle e^2\rangle=0.47^2$ is the total noise estimated in
the VIRMOS-DESCART survey and here we take it as what would be
expected by the CFHT Legacy Survey.  It includes the dispersion of
the galaxy intrinsic ellipticity, PSF correction noise and photon
shot noise. $\langle N_{\rm pixel} \rangle$ is the mean number of
galaxies in each pixel. For VIRMOS, the number density of observed
galaxies $n_g\simeq 26/{\rm arc min}^2$, then $\langle N_{\rm
pixel} \rangle=n_g[(\theta_\kappa/1^{'})/N]^2$, where $N=2048$ is
the number of grids by which we store 2D maps and the field of
view $\theta_\kappa$ is in units of arc min. The factor of 2
arises from the fact that the shear field has two degrees of
freedom ($\gamma_1,\gamma_2$), where the definition of $\langle
e^2\rangle$ sums over both. We use this as our best guess for the
CFHT Legacy Survey noise. The maps we obtained through the method
described above are non-periodic after the projection. In order to
eliminate edge effects,  we crop each smoothed map by a factor of
$10\%$ in the margins of each $\kappa$ map for model with
$\Omega_m=0.2$. For comparison, the size of maps for other models
is the same as that of $\Omega_m=0.2$.

\section{The optimal filter}

Our goal is to find the optimal filter for constraining $\Omega_m$ by
the non-Gaussianity of weak lensing. The non-Gaussianity of weak
lensing for a clean map is quantified by the skewness $S_3$
\begin{equation}
\label{eqn:cleans3}
S_3(\theta_f)=\frac{\langle \kappa^3 \rangle}{\langle \kappa^2\rangle^2},
\end{equation}
where $\theta_f$ is the characteristic radius of the filter
function.
When noise is present, the definition of skewness can be
modified to be (\cite{2000ApJ...537....1W})
\begin{equation}
\label{eqn:dirtys3}
S_3(\theta_f)=\frac{\langle \kappa_{\rm S+N}^3 \rangle}
{[\langle \kappa_{\rm S+N}^2\rangle^2-\kappa_N^2]^2}.
\end{equation}
The subscript $S$ indicates a weak lensing signal while $N$ denotes  random
noise. Since $\langle \kappa_{\rm S+N}^3 \rangle=\langle \kappa_{\rm S}^3
\rangle$, $\langle \kappa_{\rm S+N}^2 \rangle-\langle \kappa_{\rm N}^2
\rangle=\langle \kappa_{\rm S}^2 \rangle$, thus $S_3$ defined by
Eq.\ref{eqn:dirtys3}
is statistically equivalent to the one defined by Eq.\ref{eqn:cleans3},
and the presence of noise has only residual effects on the dispersion of $S_3$.

The skewness $S_3$ is a function of cosmological density
parameters $\Omega_m$ and the filter function. The noise
introduces a large dispersion in $S_3$ and also smears its
intrinsic dependence on cosmological parameters. Filtering on a
large scale reduces this noise, but also reduces the intrinsic
skewness, and increases sample variance.  Our goal is to find the
optimal smoothing scale. Different filters also have different
scale dependence,. The general form of this filter is hard to
find, so we will employ five parametrized classes of filters in
this paper. They are the top-hat (hereafter,TH), Gaussian (GS),
aperture (AP), compensated Gaussian (CG) and Wiener (WN) filters,
respectively. TH is normalized to have a  sum unity in the 2D
window function map, and GS is defined as $W(\theta)=(1/2\pi
\theta_f^2)\exp{(-\frac{\theta^2}{2\theta_f^2})}$ which is
normalized by the same as TH. AP is defined as
$W(\theta)=\frac{9}{\pi}(\frac{1}{\theta_f})^2[1-(\frac{\theta}{\theta_f})^2]
[1/3-(\frac{\theta}{\theta_f})^2]$ and zero for $\theta>\theta_f$,
which has zero mean. The CG filter is written as
$W(\theta)=\frac{1}{2\pi
\theta_f^2}\left(1-\frac{\theta^2}{2\theta_f^2}\right)
\exp\left(\frac{-\theta^2}{2 \theta_f^2}+1\right)$, which holds
zero area, and is normalized to have a peak amplitude of unity in
Fourier space. This has the feature that it will only damp modes,
and never amplify. Many analytic integrals for CG can be evaluated
analytically \citep{2002ApJ...568...20C}. WN is defined in Fourier
space by $W(l)=\frac{C_s(l)}{C_s(l)+C_n}$, where $C_s(l)$ is the
angular power spectrum of the signal $\kappa$,
 while $C_n=\frac{4\pi\sigma_N^2f_{sky}}{N^2}$ is that for noise power,
and $f_{sky}=\pi(\frac{\theta_\kappa}{360})^2$
is the fractional sky coverage of each map.

Given these filters, one can measure $S_3$ and its dispersion
$\Delta S_3$, which are all functions of $\Omega_m$. $\Delta S_3$
causes the inferred $\Omega_m$ to differ from its true value by a
change of $\Delta \Omega_m$. For each class of filter, there
exists an optimal filter radius $\theta_f$ to minimize $\Delta
\Omega_m$. Comparing the minimum $\Delta \Omega_m$ of each class
of filter, one can then find the optimal one.

\begin{figure}
\epsscale{1.}
\plotone{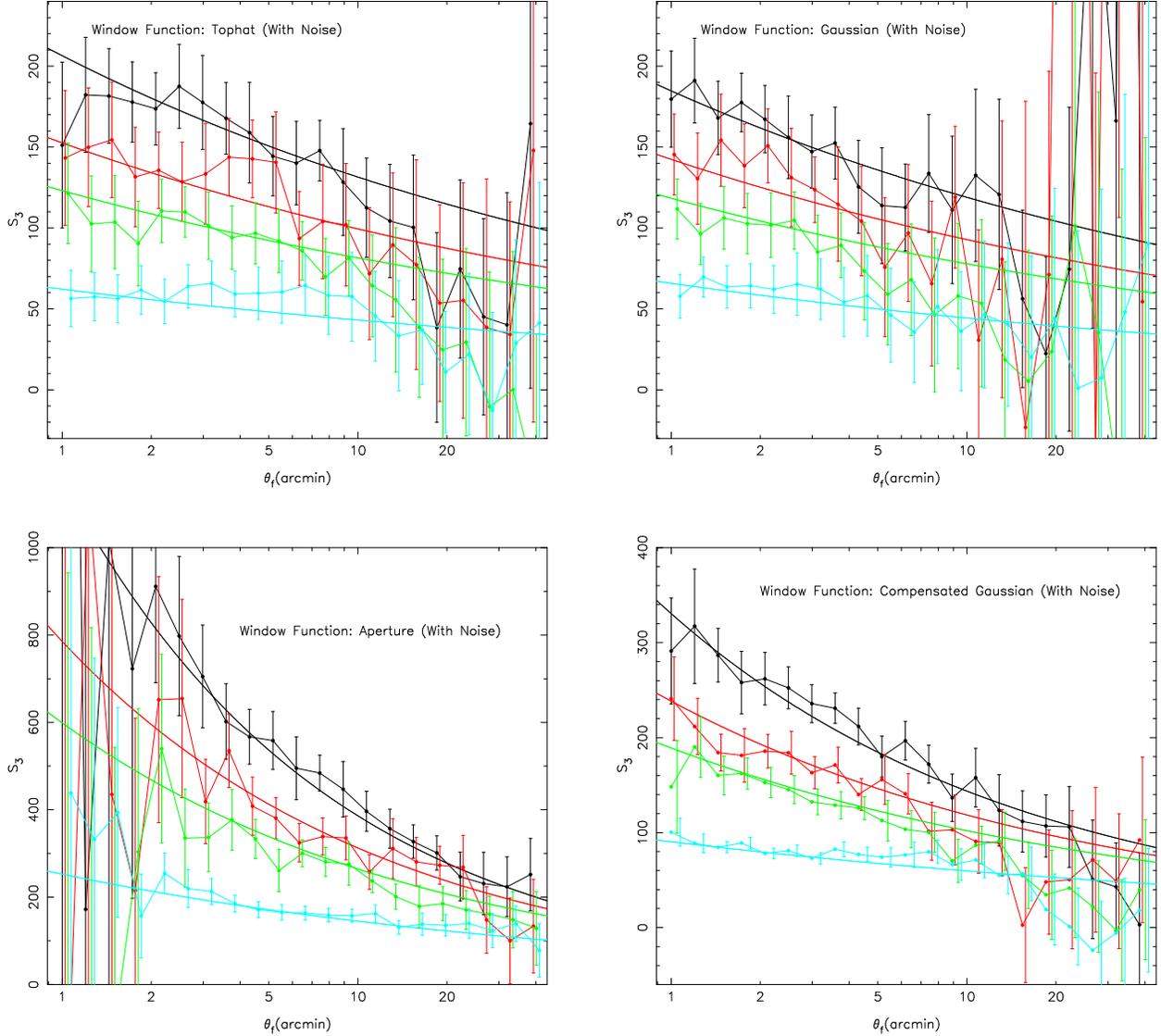}
\caption{Skewness $S_3$ as a function of a smoothing scale $\theta_f$ for all window functions
 except for Wiener with cosmological models $\Omega_m=0.2$ (black line), $\Omega_m=0.3$
(red line), $\Omega_m=0.4$(green line) and $\Omega_m=1$ (blue
line) respectively. The standard deviation of $S_3$ is taken to be
one fourth of original predicted value from simulation, because
the CFHT Legacy Survey will observe 160 square degrees, which is
about 24 times larger than the simulated area. The smoother lines
correspond to the best fit to $S_3$ for each model. For the
purpose of discerning the error bars of different models, the
plots of all models except for $\Omega_m=0.2$ are shifted in the
direction of right. The simulated convergence $\kappa$ has added
random noise.} \label{fig:s34}
\end{figure}

\begin{figure}
\epsscale{1.}
\plotone{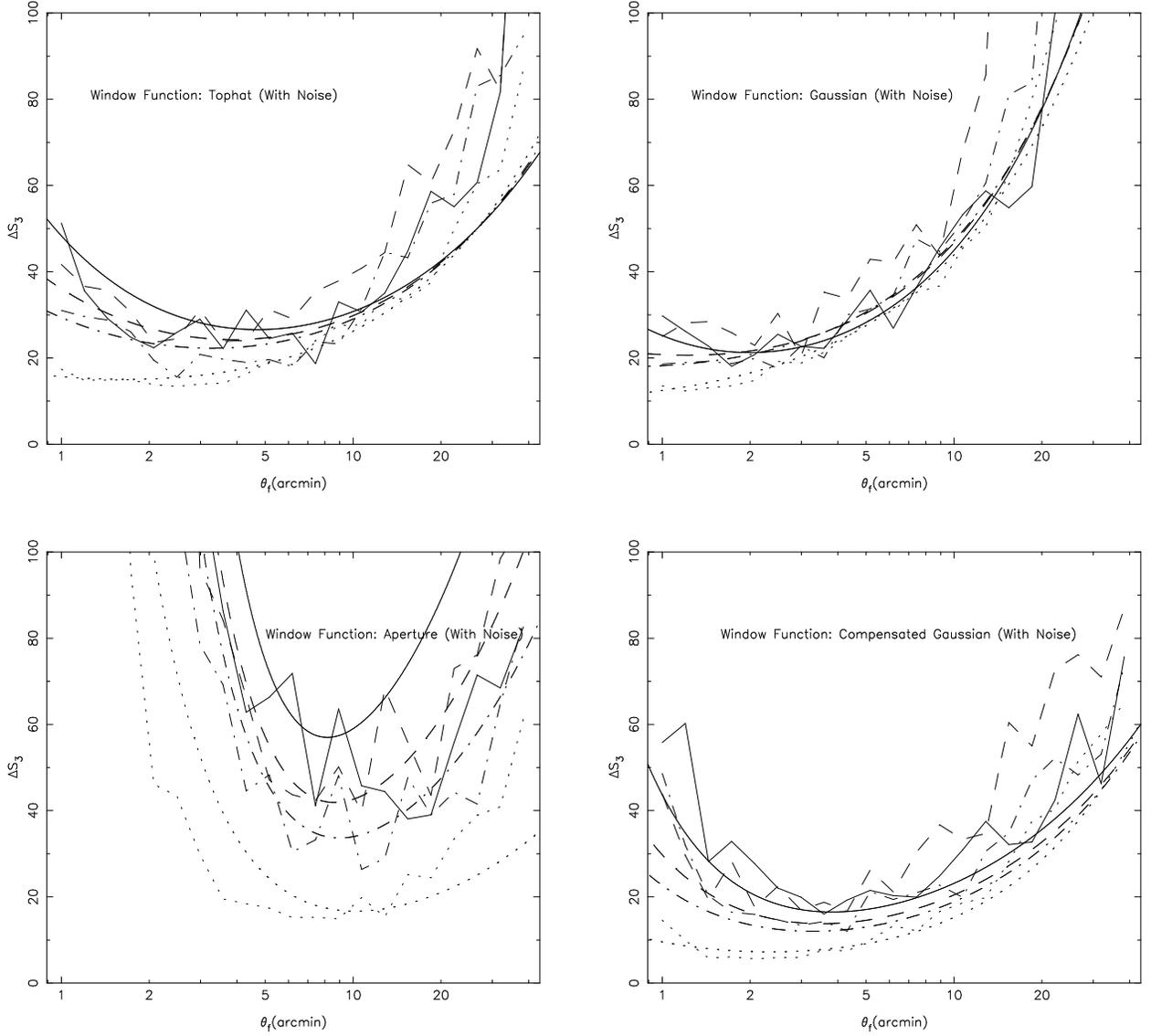}
\caption{$\Delta S_3$ as a function of a smoothing scale $\theta_f$ for all window functions
 except for Wiener with cosmological models $\Omega_m=0.2$ (solid line), $\Omega_m=0.3$
(dashed line), $\Omega_m=0.4$(dash-dotted line) and $\Omega_m=1$ (dotted line) respectively.
The smoother lines correspond to the best fit to $\Delta S_3$ for each model.}
\label{fig:dts34}
\end{figure}

\begin{figure}
\epsscale{1.}
\plotone{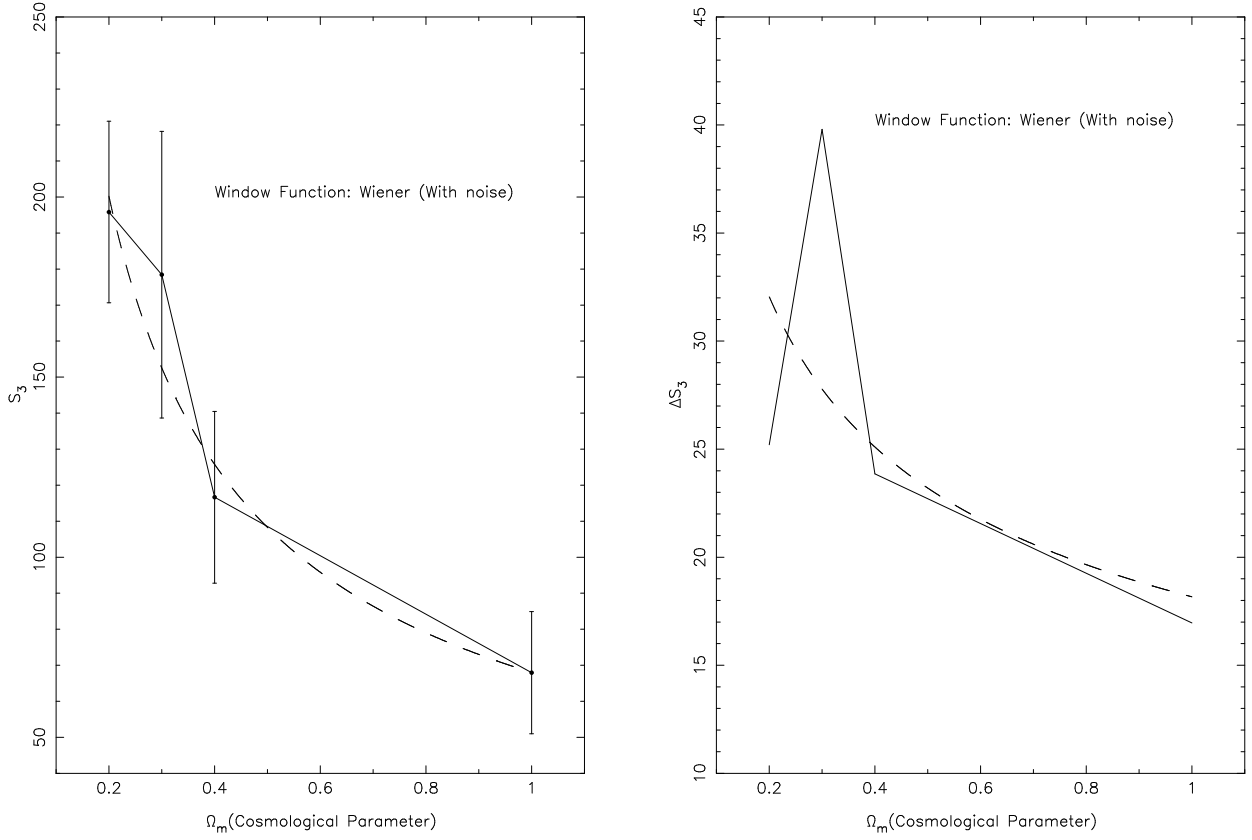}
\caption{Skewness $S_3$ (left panel) and standard deviation $\Delta S_3$ (right panel)
 as a function of cosmological parameter $\Omega_m$ for Wiener window function.
The dashed lines indicate the best fit to $S_3$ and $\Delta S_3$ respectively.
The error bars represent the same significance as Fig.\ref{fig:s34}.
The random noise is added to the simulated convergence $\kappa$.}
\label{fig:dts3wn}
\end{figure}

From simulated maps, we first calculate the skewness $S_3$ and its
standard deviation $\Delta S_3$ with different filter radius for
all kinds of cosmological models. The CFHT Legacy Survey will
observe 160 square degrees, which is about 24 times larger than
the simulated area. We conservatively decreased the error we
obtained in the field-to-field variations by a factor of about 4
to estimate the sensitivity for the Legacy Survey. Therefore, the
standard deviation of $S_3$ throughout this paper is taken to be
one fourth of original predicted value from simulation. The
skewness $S_3$ and its standard deviation $\Delta S_3$ of TH, GS,
AP and CG window functions are shown in Fig.\ref{fig:s34} and
\ref{fig:dts34}, while that of WN filter are plotted in
Fig.\ref{fig:dts3wn}. From Fig.\ref{fig:s34} and \ref{fig:dts3wn},
it is shown that the expected $S_3$ decrease with the cosmological
density parameter $\Omega_m$ for all of filters, which is in
consistent with that predicted by perturbation theory at large $\theta_f$
\citep{1998A&A...331..829G}  and non-linear
perturbation theory \citep{1999ApJ...519L...9H,2001MNRAS.322..918V}. A fixed
fluctuation amplitude measured by weak lensing is a smaller
fractional fluctuation in a higher $\Omega_m$ universe, and thus
less non-linear and less non-Gaussian. $S_3$ also decreases with
filter scale $\theta_f$, as one would expect from the central
limit theorem when smoothing over more independent patches to
converge to a Gaussian distribution.  Our dependence of $S_3$
agrees qualitatively with \citet{2001MNRAS.322..918V}, where
the detailed normalization depends on details of the redshift 
distribution.  Estimates of $S_3$ are possible analytically.
To optimize its measurement, we also need its standard deviation,
which is related to a six point function.  This is difficult
to compute analytically.

The fit for
$S_3$ in Fig.\ref{fig:s34} fails badly on not only small scale but
large scales of more than 10 arc minutes. This is due to larger
standard deviation of $S_3$ as shown in Fig.\ref{fig:dts34} on
both of the two scales. It is also apparent from
Fig.\ref{fig:dts34} that there exists an optimal filtering scale
$\theta_f$ corresponding which minimizes $\Delta S_3$ (except for
the WN filter that does not depend on filter radius). This is due
to the trade-off between noise on small scales and sample variance
on large scales. $S_3$ as a function of filter radius $\theta_f$
for the TH filter in Fig.\ref{fig:s34} is in rough agreement with
that of \cite{2000ApJ...537....1W} where cosmological model is
specified to $\Omega_m=0.3$, but it differs from
\cite{2000ApJ...530..547J}.  We do not understand the behaviour
of the Gaussian window for large $\Omega$ at small angular scale, where
a smoothing scale smaller than our resolution seems to be prefered.
But we do not dwell on this since the Gaussian window is not observable
on a shear map.

From the above analyses, it is clear that $S_3$ and $\Delta S_3$
depend not only on the smoothing scale $\theta_f$ but also the
cosmological parameter $\Omega_m$ for all window functions except
for the WN filter. In real observations of weak lensing, one must
evaluate the uncertainty in $\Omega_m$ for a given observed $S_3$
and $\Delta S_3$ to discriminate between cosmological models. One
needs to invert the relation $S_3=S_3(\Omega_m,\theta_f)$ to
obtain $\Omega_m=\Omega_m(S_3,\theta_f)$ and estimate the
uncertainty of inferred $\Omega_m$ by
\begin{equation}
\label{eqn:deltom}
\Delta \Omega_m(S_3(\Omega_m,\theta_f),\theta_f)=
\Omega_m(S_3(\Omega_m,\theta_f),\theta_f)-\Omega_m(S_3(\Omega_m,\theta_f)+
\Delta S_3(\Omega_m,\theta_f),\theta_f).
\end{equation}
Because of the irregularity of the data points, the inversion is noisy
and may introduce unrealistic artifacts. To overcome this problem, we
first fit $S_3$ and $\Delta S_3$ by a combination of power laws of $\Omega_m$
and $\theta_f$ in the presence of noise, respectively
\begin{equation}
\label{eqn:s3omth}
S_3(\Omega_m,\theta_f)=A(\Omega_m)\theta_f^{B(\Omega_m)},
\end{equation}
and
\begin{equation}
\label{eqn:delts3omth}
\Delta S_3(\Omega_m,\theta_f)=A_S(\Omega_m)\theta_f^{B_S(\Omega_m)}+
A_N(\Omega_m)\theta_f^{B_N(\Omega_m)},
\end{equation}
where $A(\Omega_m)=A_1\Omega_m^{B_1}$,
$B(\Omega_m)=A_2\Omega_m^{B_2}$,
$A_S(\Omega_m)=A_{S1}\Omega_m^{B_{S1}}$,
$B_S(\Omega_m)=A_{S2}\Omega_m^{B_{S2}}$,
$A_N(\Omega_m)=A_{N1}\Omega_m^{B_{N1}}$ and
$B_N(\Omega_m)=A_{N2}\Omega_m^{B_{N2}}$ respectively. The two
terms of $\Delta S_3(\Omega_m,\theta_f)$ represent two sources of
the dispersion of $S_3$: the intrinsic dispersion of signal and
that caused by noise. Noise dominates at small smoothing scale, so
we expect that $B_N(\Omega_m)<0$, while we expect
$B_S(\Omega_m)>0$ because of the large smoothing behavior of
$\Delta S_3$ from signal(Fig.\ref{fig:dts34}). For the Wiener
function,  we can just parameterize the dependence of skewness
$S_3$ and its standard deviation $\Delta S_3$ on $\Omega_m$ as
$S_3(\Omega_m)=A_1\Omega_m^{B_1}$ and $\Delta
S_3(\Omega_m)=A_2\Omega_m^{B_2}$ respectively, because it is due
to its independence of smoothing radius.

\begin{deluxetable}{rrrrrr}
\tablecolumns{6}
\tablewidth{0pc}
\tablecaption{Best Fit to $S_3$ and $\Delta S_3$ by Eq.(\ref{eqn:s3omth}) and
Eq.(\ref{eqn:delts3omth})}
\tablehead{
\colhead{} & \colhead{Top-hat} & \colhead{Gaussian} & \colhead{Aperture} &
\colhead{Compensated Gaussian}}
\startdata
$A_1$ & 62.16$\pm$1.21 & 65.79$\pm$1.11 & 252.74$\pm$4.01 & 90.33$\pm$0.69\\
$B_1$ & -0.75$\pm$0.02 & -0.64$\pm$0.01 & -0.94$\pm$0.02 & -0.81$\pm$0.01\\
$A_2$ & -0.16$\pm$0.03 & -0.17$\pm$0.02 & -0.24$\pm$0.01 & -0.18$\pm$0.01\\
$B_2$ & -0.13$\pm$0.11 & -0.07$\pm$0.09 & -0.42$\pm$0.02 & -0.43$\pm$0.03\\
$A_{S1}$ & 3.47$\pm$0.04 & 5.43$\pm$0.08 & 2.20$\pm$0.17 & 1.47$\pm$0.07\\
$B_{S1}$ & -0.25$\pm$0.02 & -0.07$\pm$0.03 & -0.79$\pm$0.10 & -0.71$\pm$0.05\\
$A_{S2}$ & 0.79$\pm$0.01 & 0.84$\pm$0.01 & 0.73$\pm$0.03 & 0.98$\pm$0.02\\
$B_{S2}$ & 0.10$\pm$0.02 & -0.002$\pm$0.02 & -0.06$\pm$0.04 & 0.23$\pm$0.03\\
$A_{N1}$ & 12.18$\pm$1.75 & 7.02$\pm$0.86 & 380.75$\pm$108.86 & 8.11$\pm$1.60\\
$B_{N1}$ & -0.79$\pm$0.13 & -0.62$\pm$0.12 & -1.05$\pm$0.23 & -0.98$\pm$0.17\\
$A_{N2}$ & -0.42$\pm$0.09 & -0.05$\pm$0.05 & -1.90$\pm$0.15 & -0.87$\pm$0.17\\
$B_{N2}$ & -0.43$\pm$0.18 & -1.78$\pm$0.45 & -0.14$ \pm$0.07 & -0.35$\pm$0.14\\
\label{table:fit}
\enddata
\end{deluxetable}

\begin{figure}
\epsscale{1.}
\plotone{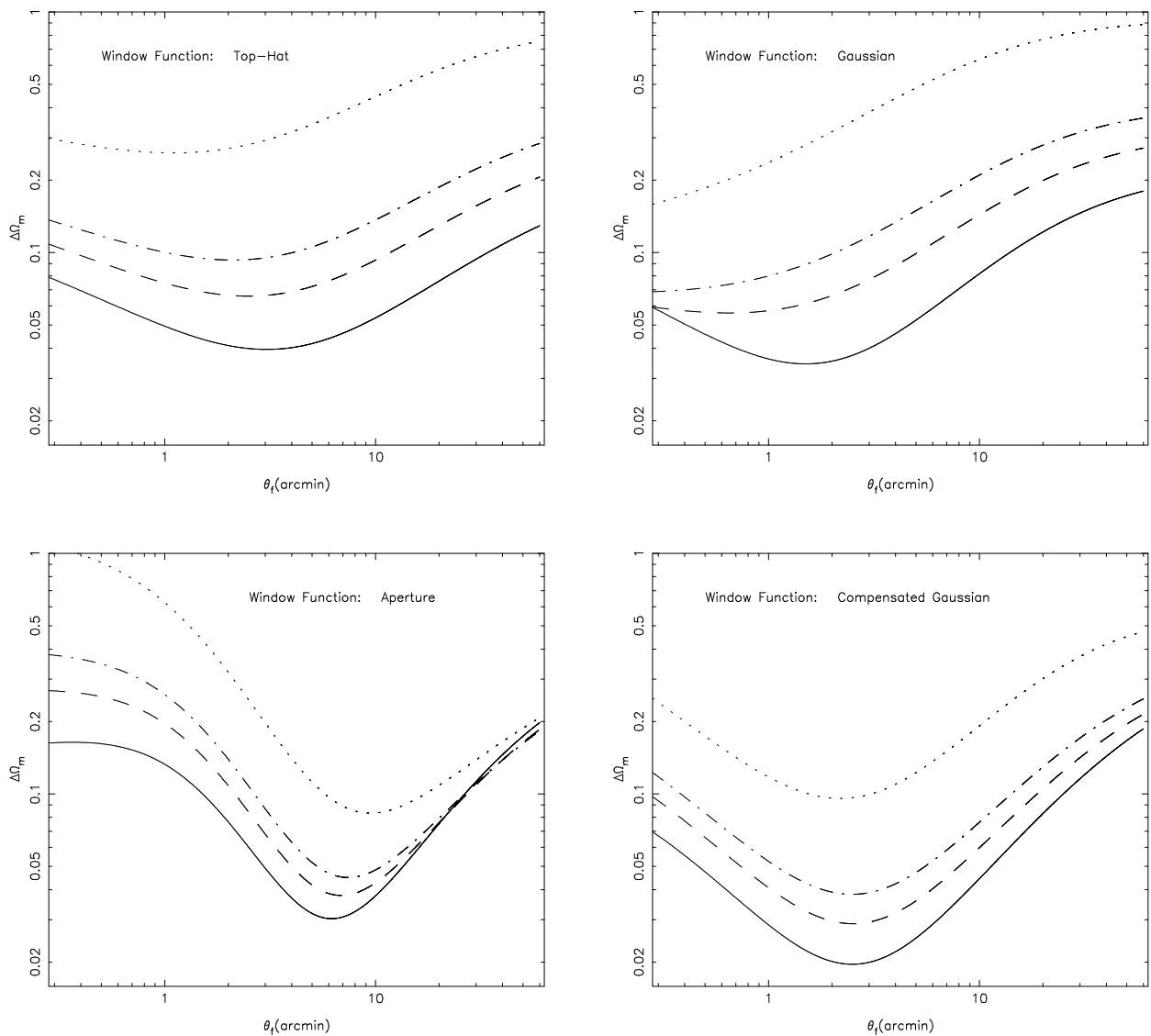}
\caption{
$\Delta \Omega_m$ as a function of a smoothing scale $\theta_f$ for all except for Wiener
window function with cosmological models $\Omega_m=0.2$ (solid line), $\Omega_m=0.3$
(dashed line), $\Omega_m=0.4$(dash-dotted line) and
$\Omega_m=1$ (dotted line) respectively. The simulated convergence $\kappa$ is
added with random noise.}
\label{fig:deltom4}
\end{figure}

\begin{figure}
\epsscale{1.}
\plotone{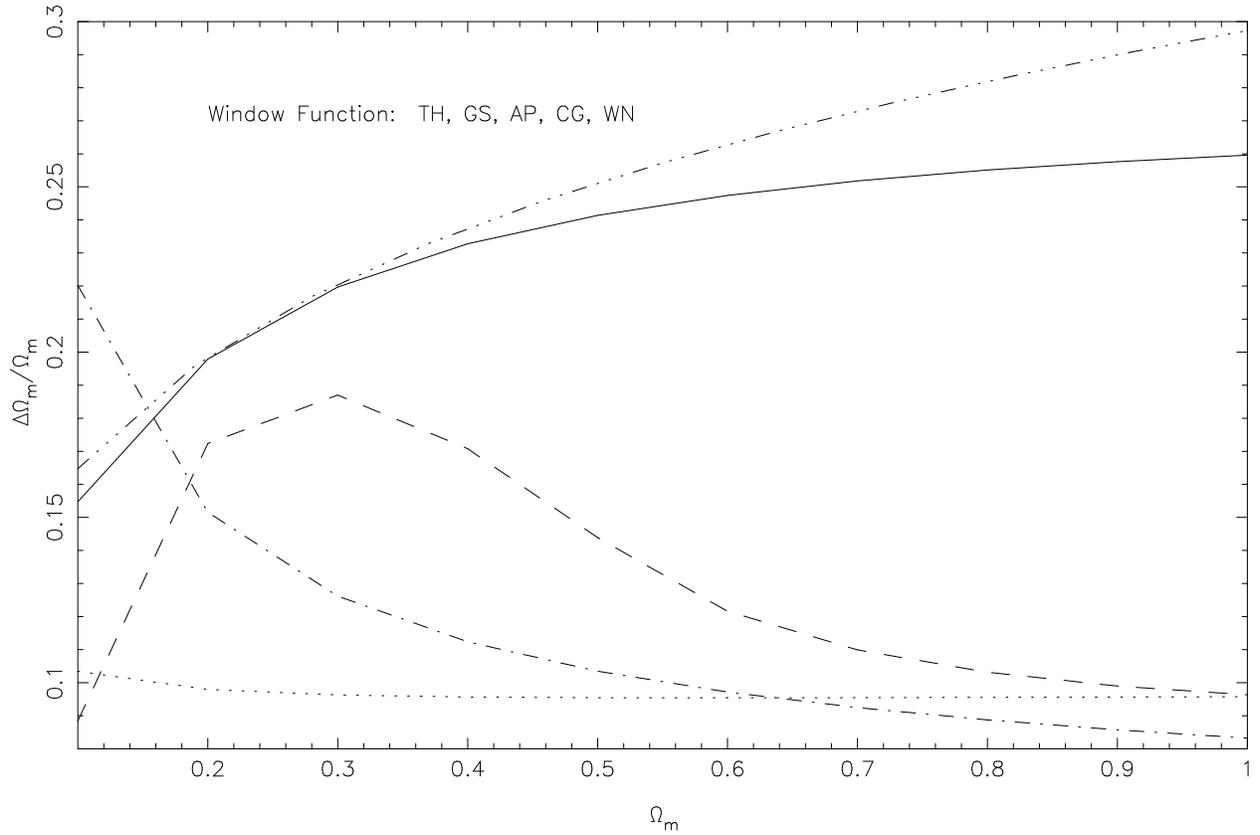}
\caption{The relative uncertainty $\Delta \Omega_m/\Omega_m$ as
a function of $\Omega_m$ smoothed at the optimal filter radius for Top-hat(solid line),
Gaussian(dashed line), Aperture(dash-dotted line),
Compensated Gaussian(dotted line) and Wiener(dash-tridotted line) filters
respectively.}
\label{fig:deltom5}
\end{figure}

We fit the relation of $S_3$ and $\Delta S_3$ with a function of
$\Omega_m,\theta_f$ by Eqs.(\ref{eqn:s3omth}) and
(\ref{eqn:delts3omth}) for all filters except for WN. Their best
fit coefficients $A_{S1},B_{S1},A_{S2},B_{S2},A_{N1},B_{N1}$,
$A_{N2},B_{N2}$  are listed in Table \ref{table:fit} respectively,
the best fit relations of which are also plotted with smoother
lines in Fig.\ref{fig:s34} and \ref{fig:dts34}. For the WN filter, the
best fit coefficients $A_1=68.03\pm2.37$, $B_1=-0.67\pm0.03$,
$A_2=18.17\pm6.13$ and $B_2=-0.35\pm0.26$, and the dashed lines in
Fig.\ref{fig:dts3wn} show the best fit to $S_3$ and $\Delta S_3$.
In the fit of skewness $S_3$, we weighted using the standard
deviation of the skewness. Using these best fit coefficients, we
can calculate the $\Delta \Omega_m$ with the function of
$\Omega_m$ and $\theta_f$, which are shown in
Fig.\ref{fig:deltom4} and \ref{fig:deltom5}. As expected, we find
in Fig.\ref{fig:deltom4} that there does exist an optimal
smoothing scale for each class of filter (except for the WN filter) that
has a minimum error for the inferred $\Omega_m$. This optimal
smoothing scale has only a weak dependence on cosmology except for
GS filter. The minimum $\Delta \Omega_m$ decreases toward lower
$\Omega_m$. Due to the $S_3\propto \Omega_m^{\sim -0.8}$ behavior
(Table \ref{table:fit}), at low $\Omega_m$, a small change of
$\Omega_m$ results in a large change of $S_3$. But $\Delta S_3$
does not have such strong $\Omega_m$ dependence, thus the
resulting error in $\Omega_m$ decreases toward lower $\Omega_m$.
In addition, we show in Fig.\ref{fig:deltom5} the relative
uncertainty $\Delta \Omega_m/\Omega_m$ as a function of $\Omega_m$
smoothed at the optimal filter radius for all of filters
respectively. It is shown that the relative uncertainty $\Delta
\Omega_m/\Omega_m$ for the Compensated Gaussian filter almost stays
constant with $\Omega_m=0.1$, and takes the smallest value
in the range from $\Omega_m=0.2$ to 0.6 of interest compared with
that of other filters. By comparing the minimum of $\Delta
\Omega_m$ for each filter class, we then conclude that the
compensated Gaussian filter to be the optimal filter for all
cosmologies. The relative uncertainty $\Delta \Omega_m/\Omega_m$
at the optimal filter scale for this filter is nearly independent
of $\Omega_m$, and the
 corresponding optimal filter scale is about 2.5 arc minutes.

\section{Conclusions}

We have studied the power of weak lensing surveys to measure the matter density
of the universe without relying on any exterior data sets.  We found
that the CFHT Legacy Survey can measure a fractional accuracy in $\Omega_m$ of
10\%, which is competitive with global joint analyses, but bypasses
a large number of cross calibration uncertainties.

We ran a series of high resolution N-body simulation to study
statistical skewness properties of weak lensing by large-scale
structure in the universe with a range of cosmological matter
density parameters. We added noise due to intrinsic ellipticity of
background faint galaxies to the simulated $\kappa$ fields and
smoothed it using different filters with a range of smoothing
radii.  We calculated the skewness $S_3$ of the smoothed $\kappa$
field with added Gaussian noise and predicted the uncertainty
$\Delta \Omega_m$ for the cosmological mass density parameter for
a given $S_3$ and smoothing radius $\theta_f$. We examined the
relative discriminating power of different window functions for
distinguishing cosmological models in the upcoming CFHT Legacy
Survey.  Except for the Wiener filter, we found the optimal
smoothing radius for all of four window functions that minimizes
$\Delta \Omega_m$.  This optimal smoothing scale has only a weak
dependence on cosmology.  The compensated Gaussian function was
the optimal filter for measuring $\Omega_m$ from skewness.  The
relative uncertainty $\Delta \Omega_m/\Omega_m$ smoothed at the
optimal filter radius for Compensated Gaussian filter is about
10\%. 

To overcome the irregularity of the simulated $S_3$ and $\Delta
S_3$, we have fitted their smoothing scale and cosmology
dependence with some phenomenological power laws. One could derive
these relations analytically using perturbation theory following
the theoretical work of \citet{1997A&A...322....1B}. But since
skewness is intrinsically non-linear, such a perturbation approach
has to be tested against simulations. In fact, in our work based
on simulations, the optimal filter radius is a few arc minutes or
$\sim 1$ Mpc/h, which lies in strongly non-linear regime, where
perturbation theory breaks down \citep{1998A&A...331..829G}. In
the non-linear regime, a semi analytical model,
hyperextended perturbation theory (hereafter HEPT)
\citep{1999ApJ...520...35S}, which applies at the highly non-linear regime, and
a fitting formula to interpolate between the quasi-linear and highly non-linear
regime \citep{2001MNRAS.325.1312S}, have been applied to predict $S_3$
\citep{1999ApJ...519L...9H,2001MNRAS.322..918V}. Since these models reply on
simulations for calibration, they by no means can produce better
result than simulations.  Furthermore, to calculate the lensing
$\Delta S_3$ analytically, one has to know the $S_6$ of the
density field, which can be predicted by HEPT but has not been
tested against simulations \citep{1999ApJ...520...35S}. So we'd
rather using our fitting formula approach instead of adopting
these analytical results.

We thank an anonymous referee for several helpful suggestions on
the manuscript. T.J.Zhang would like to thank CITA for its
hospitality during his visit. The research was supported in part
by NSERC and computational resources of the CFI funded
infrastructure. P.J.Zhang thanks the department of astronomy \&
astrophysics and CITA of University of Toronto where part of the work was done.
P.J.Zhang is supported by the DOE and the NASA grant NAG 5-10842
at Fermilab. 

\bibliography{penbib}
\bibliographystyle{apj}

\appendix

\end{document}